# Design and implementation of the NaI (Tl)/CsI (Na) detectors output signal generator[*]

ZHOU Xu(周旭)[1,2] LIU Cong-Zhan(刘聪展)[1;1)] ZHAO Jian-Ling(赵建领)[1] ZHANG Fei(张飞)[1] ZHANG Yi-Fei(张翼飞)[1] LI Zheng-Wei(李正伟)[1,2] ZHANG Shuo(张硕)[1,2] LI Xu-Fang(李旭芳)[1] LU Xue-Feng(路雪峰)[1] XU Zhen-Ling(许振玲)[1] LU Fang-Jun(卢方军)[1]

[1] Institute of High Energy Physics, Chinese Academy of Sciences, Beijing 100049, China
[2] University of Chinese Academy of Sciences, Beijing 100049, China

**Abstract**:We designed and implemented a signal generator that can simulate the output of the NaI (Tl)/CsI (Na) detectors' pre-amplifier onboard the Hard X-ray Modulation Telescope (HXMT). Using the development of FPGA (Field Programmable Gate Array) with VHDL language and adding random constituent, we have finally produced the double exponential random pulse signal generator. The statistical distribution of signal amplitude is programmable. The occurrence time intervals of adjacent signals content negative exponential distribution statistically.

**Key words**: FPGA, M sequence, rejection technique, Gaussian distribution, signal generator

## 1 Introduction

HXMT is the first X-ray astronomy satellite in China. The High Energy X-ray Detector (HED), which consists of 18 collimated units, is the core payload on HXMT. Every unit of HED is made up of a NaI (Tl)/CsI (Na) phoswich, photomultiplier tube (PMT), high-voltage divider and pre-amplifier [1]. This paper introduces a new random pulse generator to simulate the pre-amplifier's output of an actual scintillator detector, such as the HED. The project of design, electric circuit principle design, software design of random pulse generator and pulse output circuit are introduced. The amplitude and time intervals of the signal are adjustable, and so the generator can be used as a safe and low cost signal source for the High energy Electronics System (HES) of HXMT in testing, such as vacuum, temperature, vibration and EMC (Electromagnetic Compatibility) tests [2]. The experiences we obtained are also useful for the development of other kinds of random pulse generators [3].

There are three kinds of interactions between high energy photons and materials: photoelectric effect, Compton scattering and pair production. For the HED, whose main purpose is to detect X-ray in energy band 20-250 keV, the main interaction is photoelectric effect, as illustrated in Fig. 1. But at energies higher than 100 keV, the Compton scattering effect cannot be simply neglected. Therefore, our signal generator needs to not only simulate the photoelectric effect but also the Compton scattering effect.

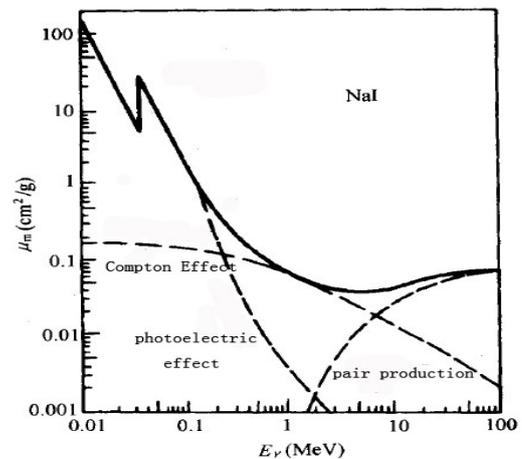

Fig. 1 The mass-absorption coefficient of NaI crystal [4].

## 2 Design of the signal generator

When an X-ray photon hits on the NaI (Tl) or CsI (Na) crystal of the HED, fluorescence light is produced, and then collected by the PMT coupled to the back of the CsI (Na) crystal. The light has a fast exponential rise followed by a slower exponential decay, and the amount of the light is proportional to the energy of the incident photon. To simulate the features of the actual

---
[*]Supported by the 973 Program (2009CB824800), NSFC (10978001 and 11003011),and the Knowledge Innovation Program of Chinese Academy of Sciences (200931111192010) .
1) E-mail: liucz@ihep.ac.cn

HED signal, the generator should be able to produce a double exponential shape wave whose expected amplitude is adjustable. Furthermore, the statistical distribution of the occurrence time intervals should follow a negative exponential function.

The function block diagram is shown in Fig.2. FPGA is the centre part of the wave generation.

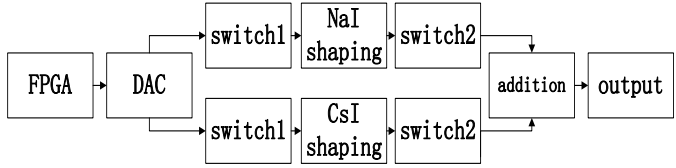

Fig. 2 Illustration of the function blocks of the signal generator

The design idea of FPGA is shown in Fig.3.

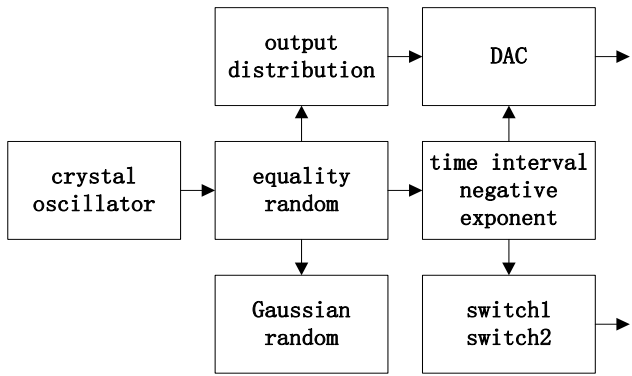

Fig. 3 Diagrammatic drawing of FPGA inside.

## 2.1 Realization of the double exponential wave

The design of the shaping circuit to simulate the signals of NaI (Tl) and CsI (Na) detectors' pre-amplifier is shown in Fig.4.

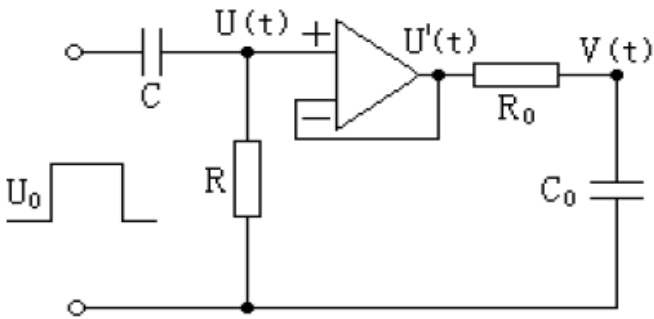

Fig. 4 The design drawing of detector signal simulation [2].

The circuit can produce a positive pulse at the rising edge of the square wave and produce a negative pulse at the falling edge. Useless negative pulse is removed by an analog switch. Positive pulse is used to generate the double exponential signal. The detailed analysis is described as follows.

For the CR loop:

$$I(t) = \frac{U(t)}{R} \quad ; \quad (1)$$

$$U(t) = U_0 - \frac{Q}{C} = U_0 - \frac{1}{C}\int_0^t I(t')dt' \quad ; \quad (2)$$

Therefore:

$$U(t) = U_0 e^{-t/RC} \quad . \quad (3)$$

For the $R_0C_0$ loop:

$$U'(t) = U(t) = U_0 e^{-t/RC} \quad ; \quad (4)$$

$$V(t) = \frac{Q_0}{C_0} = \frac{1}{C_0}\int_0^t \frac{U'(t') - V(t')}{R_0} dt' \quad ; \quad (5)$$

Therefore:

$$V(t) = U_0 \frac{RC}{RC - R_0C_0}(e^{-t/RC} - e^{-t/R_0C_0}) \quad . \quad (6)$$

As a result, the expression of PMT's output of the phoswich detector is that:

$$U^{NaI+CsI}(t) = U_0^{NaI} \frac{R^{NaI}C^{NaI}}{R^{NaI}C^{NaI} - R_0C_0}(e^{-t/R^{NaI}C^{NaI}} - e^{-t/R_0C_0})$$
$$+ U_0^{CsI} \frac{R^{CsI}C^{CsI}}{R^{CsI}C^{CsI} - R_0C_0}(e^{-t/R^{CsI}C^{CsI}} - e^{-t/R_0C_0}) \quad . \quad (7)$$

The first-part of the expression is the contribution of NaI, and the second-part accounts for the CsI signal. $U_0$ is the gain coefficient of each signal; $R_0C_0$ is the parameter (300ns, empirical value) of forming time in the circuit. It depends on the charge-discharge in the $R_0C_0$ loop. By assigning RC (RC stands for the fluorescence light decay constant of the scintillator. Here, NaI is 230ns and CsI is 630ns). In addition, rising-time, falling-time and wide are confirmed. We can produce the wave as we hope, which is shown in Fig. 5.

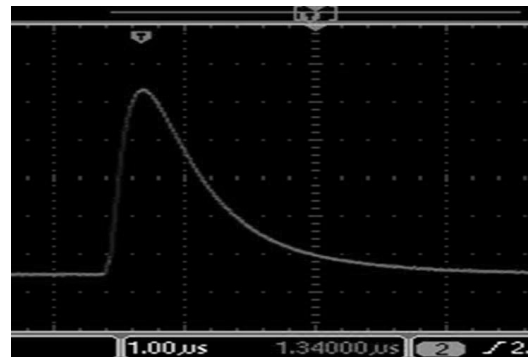

Fig. 5 The oscilloscope screenshot of the NaI wave.

In practice, we use fixed value capacitor C and adjustable resistor R, so that different decay time can be produced [5] [6].

## 2.2 The method to make the expected amplitude adjustable

The thought of M sequence is used in the random number module. The widely used M sequence is also called pseudorandom sequence or pseudorandom code, which has extensive applications in communication as an equally distributed random number generator [7]. We construct the M sequence by using the VHDL language, and collect the output of being transformed by the DAC (Digital Analog Converter) in the multichannel analyzer MCA8000A, as shown in Fig. 6.

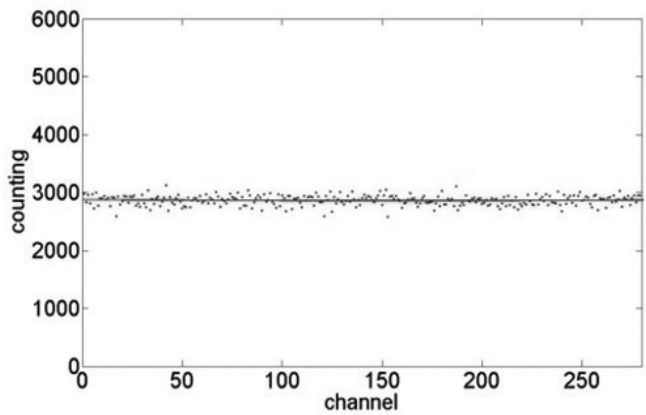

Fig. 6 Equiprobability random statistical distribution generated by the random number module. (the fitting function $f(x) = 0.01x + 2866$, RMSE: 89.96).

On the basis of central limit theory: for N equally distributed random variables, if N is large enough, the distribution of the sum is close to a Gaussian distribution. Test shows that when N is equal to or larger than 12, the result is almost perfect [3]. So in this paper we make N equals to 12. That means the sum of 12 different initial values' random series contents Gaussian distribution. We can make it in the FPGA. The output of the signal generator is transformed by the DAC, and we sample the output of the DAC by the multichannel analyzer, as shown in Fig.7.

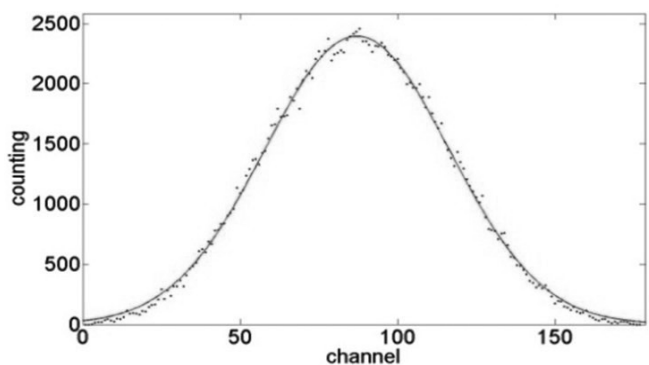

Fig. 7 Gaussian statistical distribution obtained by the M sequence method (the fitting function $f(x) = 2394*\exp(-((x-87)/41)^2)$, R-square: 0.9966).

Meanwhile, we can simulate the photoelectric effect conditions at different energies. For example, we can simulate the photoelectric effects at 60 keV, 122 keV and 250 keV at the same time. The probability of X-ray photons absorbed by the NaI (Tl) or CsI (Na) crystal at different energies is obtained by GEANT4 simulation. Table 1 listed below is the resulted approximate fraction of X-ray photons having photoelectric effect in NaI or CsI scintillation crystals.

Table 1. The probability of photoelectric interaction of X-ray photons with different energies in the NaI (Tl) or the CsI (Na) scintilators.

| Scintillator | thickness | 60keV | 122keV | 250keV |
| --- | --- | --- | --- | --- |
| NaI(Tl) | 3.5mm | 100％ | 75％ | 20％ |
| CsI(Na) | 40mm | 0％ | 25％ | 80％ |

In order to meet the proportion demands above, for an incident X-ray photon with a specific energy, we set a threshold in the code. If the equiprobability random is smaller than the threshold, the signal generator will output a NaI (Tl) signal, otherwise a CsI (Na) signal will be output. Then the energy spectrum is sampled by the multichannel analyzer, as shown in Fig.8.

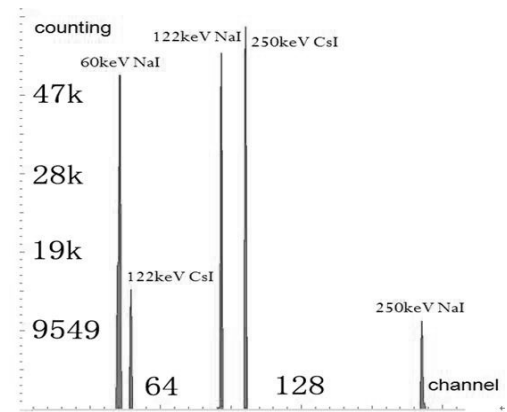

Fig. 8 The energy spectrum of the simulated NaI (Tl)/CsI (Na) detection of the 60keV, 122keV and 250keV X-ray photons.

Furthermore, based on the rejection technique, Compton scattering effect is added. The thought of the rejection technique is that we can reject the random numbers that we don't want to use and the rest of the random numbers are accepted.

The actual energy spectrum has two dimensions of information: the channel and the count number. We load the actual energy spectrum into the RAM of the FPGA. Then we generate two uncorrelated, equal-length random series A (stands for channel, value-range equals to channel) and B (stands for counts, value-range equals to counts) in the FPGA. To a given number i, we can get A[i] and B[i] from series A and B respectively. From the pre-loaded spectrum, we get C according to the value of A[i]. If B[i]is less than C then A[i] is sent to the DAC as the amplitude of the next pulse. In this way, the output has the same probability density distribution of the energy spectrum. There are two energy spectrums collected by the multichannel analyzer below, as shown in Fig.9 and Fig.10. Fig.9 is an actual energy spectrum and spectrum in Fig.10 is generated using our pulse generator.

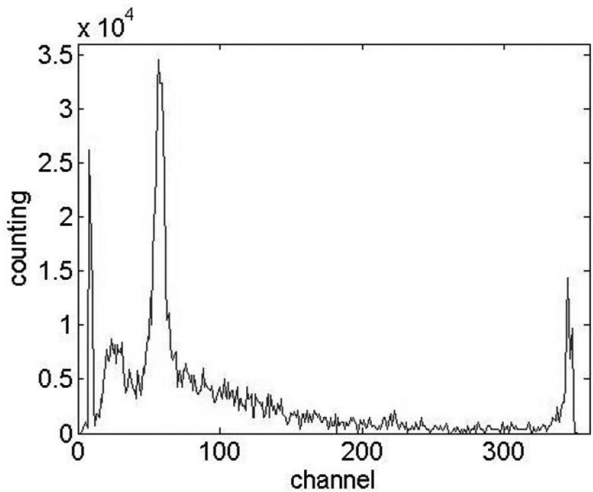

Fig. 9 The energy spectrum of [241]Am radioactive source measured in NaI and CsI scintillation crystals.

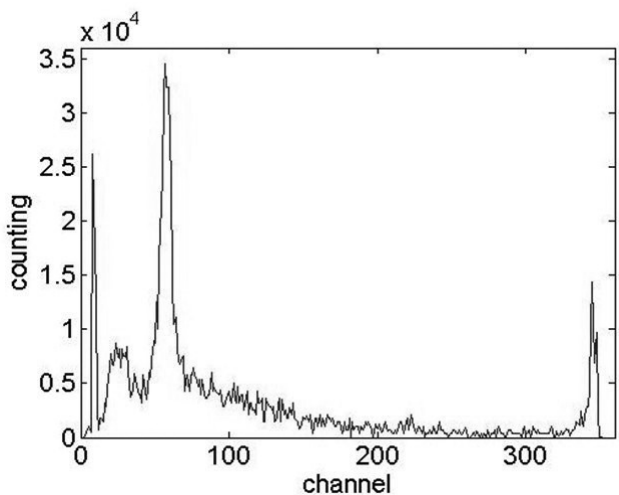

Fig. 10 The simulated energy spectrum of [241]Am radioactive source in NaI and CsI scintillation crystals.

In all, three methods for imitating energy spectrum are introduced. There is the central limit theory for the Gaussian statistical distribution, threshold comparison for the photoelectric effect and the rejection technique for the Compton scattering effect adding. Meanwhile, M sequence for equiprobability random distribution is the base of the central limit theory and the rejection technique.

## 2.3 The negative exponential distribution of the intervals between two neighboring events

We compare the 16 bits equiprobability binary random number with the specific threshold Y (we set up Y=2000) in FPGA. If the random number is smaller than Y, the result z is assigned to '1'; otherwise, z is '0'. Obviously, the probability of '1' is $p=Y/2^{16}$ in one test. If the result of the xth test is '1', and the results of the previous (x-1) test are all '0', then the probability is that：

$$f(x)=(1-p)^{(x-1)}p=\frac{p}{1-p}e^{x\ln(1-p)}. \quad (8)$$

Assigned a discrete stochastic variable X with probability mass function, because ln（1-p）<0, then X contents negative exponential distribution [2] . We output the signal if z is '1'. Obviously, the time-interval of adjacent signals is a stochastic variable with a negative exponential distribution. We sample the output of the signal generator and finally get the fitting chart as shown in Fig. 11.

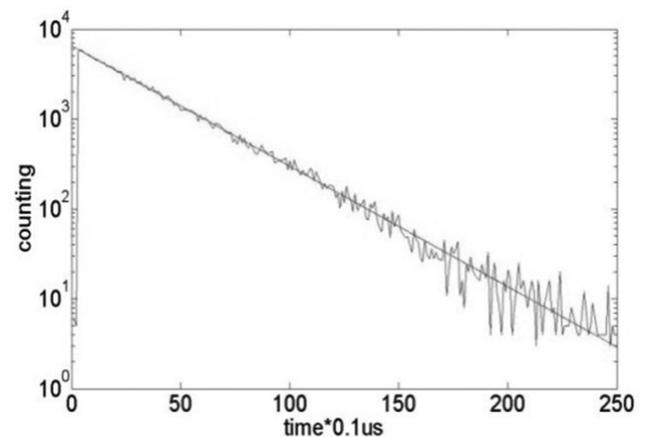

Fig. 11 The fitting chart of the signal time-interval (the fitting function f(x) = 6533*exp (-0.03x), R-square: 0.9983).

## 3 Discussion and conclusions

We developed a new type of signal generator to simulate the output of scintillator detectors' pre-amplifier. The shaping features (amplitude, rising-time, falling-time, etc.) of the signal from this generator can be adjusted easily to simulate detectors with different configurations.

By testing, the max counting rate of the signal generator is about 10000/s, and the signal generator's output amplitude is from 0.05V to 3.2V. Its output impedance is 14Ω and its quality is 60g. Meanwhile, it has the low power about 0.35W. Generating a simple cyclical signal, our signal generator (instead of the Agilent's signal generator) is collected by the multichannel analyzer. We find that the most counting is in the same channel, the proportion of the neighboring channel's counting is less than 1‰, so the generator's stability is perfect.

To imitate the output of an actual scintillator detector's pre-amplifier, it is usual to make simulation signal take on statistical distribution on time and amplitude synchronously [3]. However, it is hard to make the expected amplitude adjustable. In this paper, the rejection technique is discussed to add the Compton scattering effect. After the statistical test and performance parameter testing to output pulse, this generator truly imitates the output of an actual scintillator detector's pre-amplifier. The amplitude and occurrence time are random, the amplitude spectrum is programmable, and the occurrence time interval of adjacent signals contents a negative exponential distribution. In brief, the signal from this generator has almost the same behavior to a real detector's pre-amplifier.